\documentclass[a4paper,11pt]{article}
\usepackage{pos}
\usepackage{graphicx}
\graphicspath{ {./images/} }
\usepackage{caption}
\usepackage{subcaption}
\usepackage{svg}
\usepackage{comment}
\usepackage{hyperref}

\title{Moveable Thermometer System in ProtoDUNE}
\manuallySeparateAuthors

\author*[a]{Ranjan Dharmapalan}
\author[]{for the DUNE Collaboration}

\affiliation[a]{University of Hawai`i (Manoa),\\
  2505 Correa Rd., Honolulu, HI 96822, USA}

\emailAdd{ranjand@hawaii.edu}

\abstract{
The movable temperature profiler is a 7 m vertical array of 24 sensors that measures cryogenic temperatures with a precision of a few mK. This precision is necessary to monitor the efficiency of re-circulation and purification of liquid-argon inside large liquid-argon based neutrino detectors. Liquid argon temperature impacts electron (signal) drift velocity, flow, purity distribution and thus the overall energy calibration. The temperature profiler is motorized and moves vertically, while in the detector, and cross-calibrates neighboring sensors. The temperature offsets between each sensor cancel the effects of electromagnetic noise. This poster reports on the temperature measurements and such {\it{in-situ}} cross-calibrations at ProtoDUNE (single phase) at CERN.}

\FullConference{%
  40th International Conference on High Energy physics - ICHEP2020\\
  July 28 - August 6, 2020\\
  Prague, Czech Republic (virtual meeting)
}

\begin{document}
\maketitle

\section{Introduction}
The Deep Underground Neutrino Experiment (DUNE)\cite{duneTDR_vol1, duneTDR_vol2,duneTDR_vol3, duneTDR_vol4} will employ the world's largest Liquid Argon Time Projection Chamber (LArTPC) as its far detector. Precise monitoring of the 17 kiloton (10 kiloton fiducial mass) liquid-argon temperature will be crucial to achieve the ultimate physics goals of the experiment. ProtoDUNE-SP \cite{protodune_SP_TDR} is a 0.7 kiloton scale prototype experiment at the CERN Neutrino Platform. ProtoDUNE-SP collected charged-particle beam data and was operational for two years as a test-bed for the technologies to be employed in the eventual DUNE experiment.

\section{Dynamic Temperature Profiler}
\begin{figure}[!htb]
    \centering
    \begin{minipage}{.48\textwidth}
        \centering
        \includegraphics[width=0.95\linewidth, height=0.19\textheight]{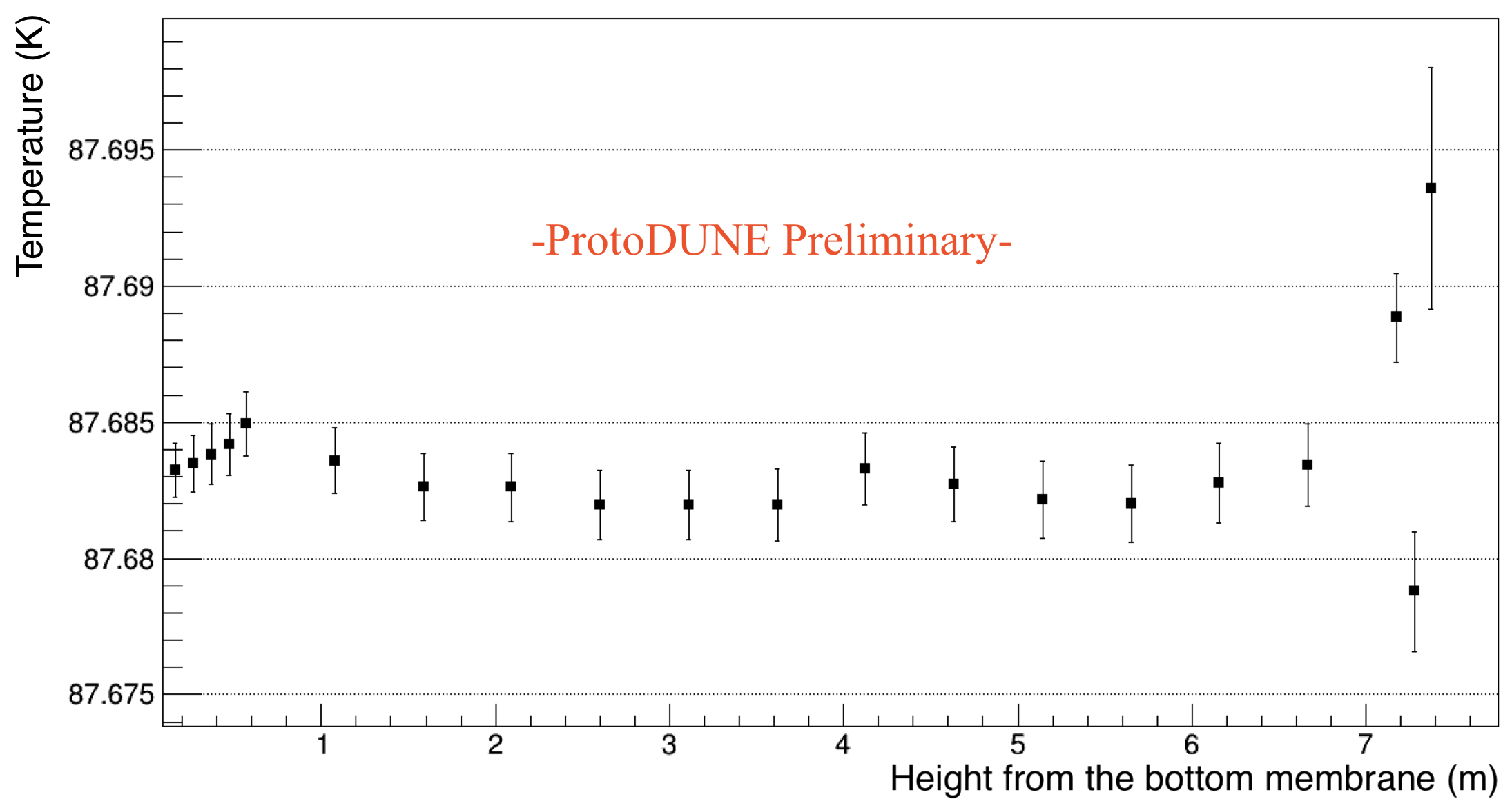}
        \caption{The dynamic temperature profiler sensor values as a function of height from the bottom of the detector. Most of the values are within a 3 mK band for sensors immersed in liquid-argon. Beyond 7 m the sensors are in the gas-liquid interface.}
        \label{fig:post_1}
    \end{minipage}%
    \hfill
    \begin{minipage}{0.48\textwidth}
        \centering
        \includegraphics[width=0.9\linewidth, height=0.17\textheight]{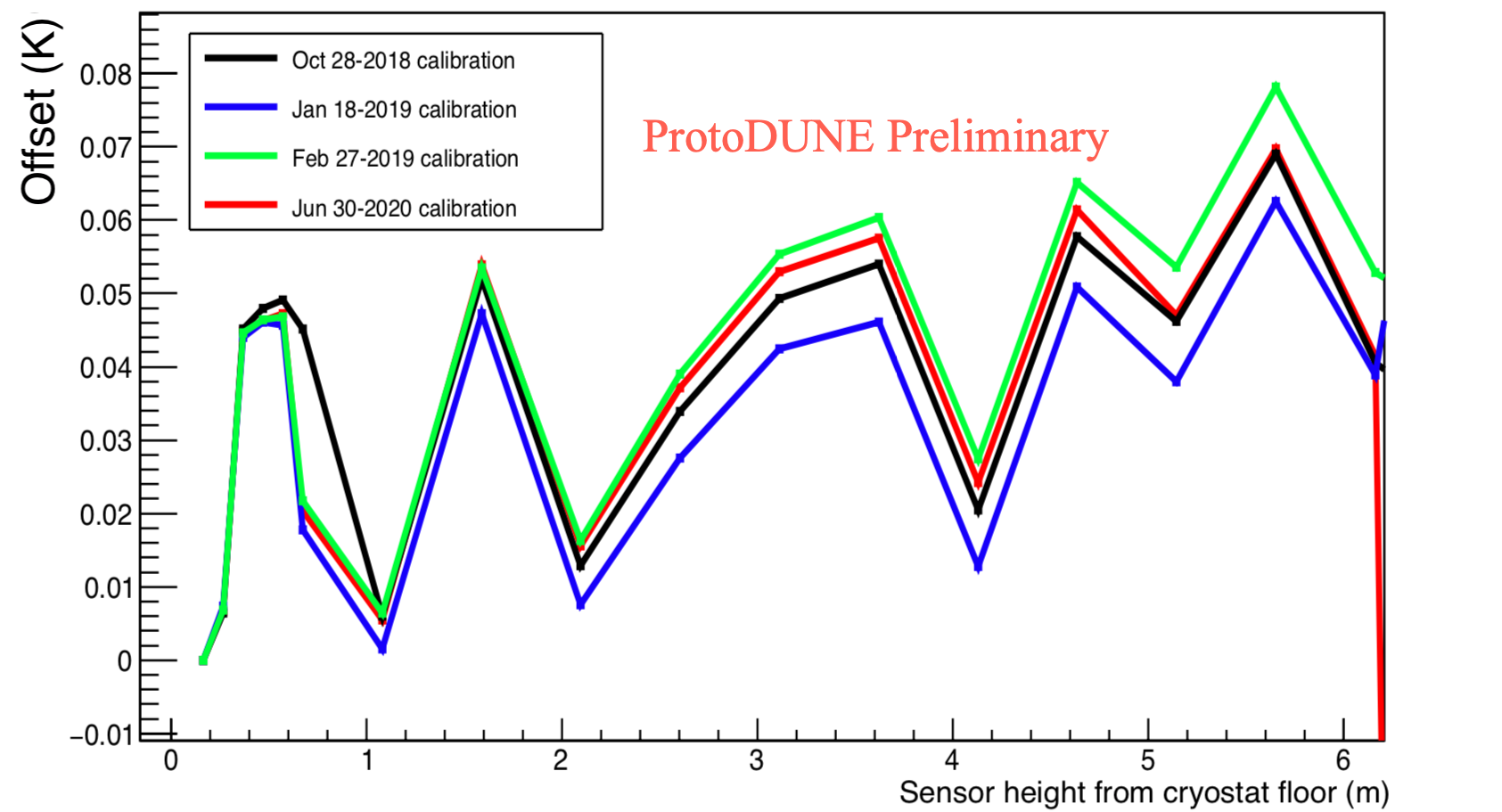}
        \caption{Comparison of the dynamic temperature profiler sensor offsets across the four calibration campaigns. The offsets are consistent and the small differences are attributed to local variations during movement of the profiler. The one outlier (October 2018 calibration of the 6th sensor from bottom) eventually converged to give consistent readings in subsequent calibration campaigns.}
        \label{fig:plot2}
    \end{minipage}
\end{figure}

The dynamic temperature profiler is a 7.5 meter vertical array of 24 sensors mounted on a stainless steel rod and accessed through a port on the downstream end of the cryostat. The top end of the rod has a rack and pinion arrangement that allows the rod to be moved up or down using a stepper motor. The sensors are platinum resistance thermometers (Lake Shore model PT-102), capable of measuring cryogenic temperatures with a precision of few millikelvin. The sensors are nominally 0.5 m apart, except near the ends, where they are clustered closer. 
The ability to move the dynamic profiler allows for cross-calibration of the sensors, as each location can be read by two sensors, thus correcting for residual parasitic resistance and electromagnetic noise and yielding a relative accuracy of 2 to 3mK (See Fig.\ref{fig:post_1}).

\section{Operation at ProtoDUNE}
\begin{figure}[!htb]
    \centering
    \begin{minipage}{.48\textwidth}
        \centering
        \includegraphics[width=0.9\linewidth, height=0.17\textheight]{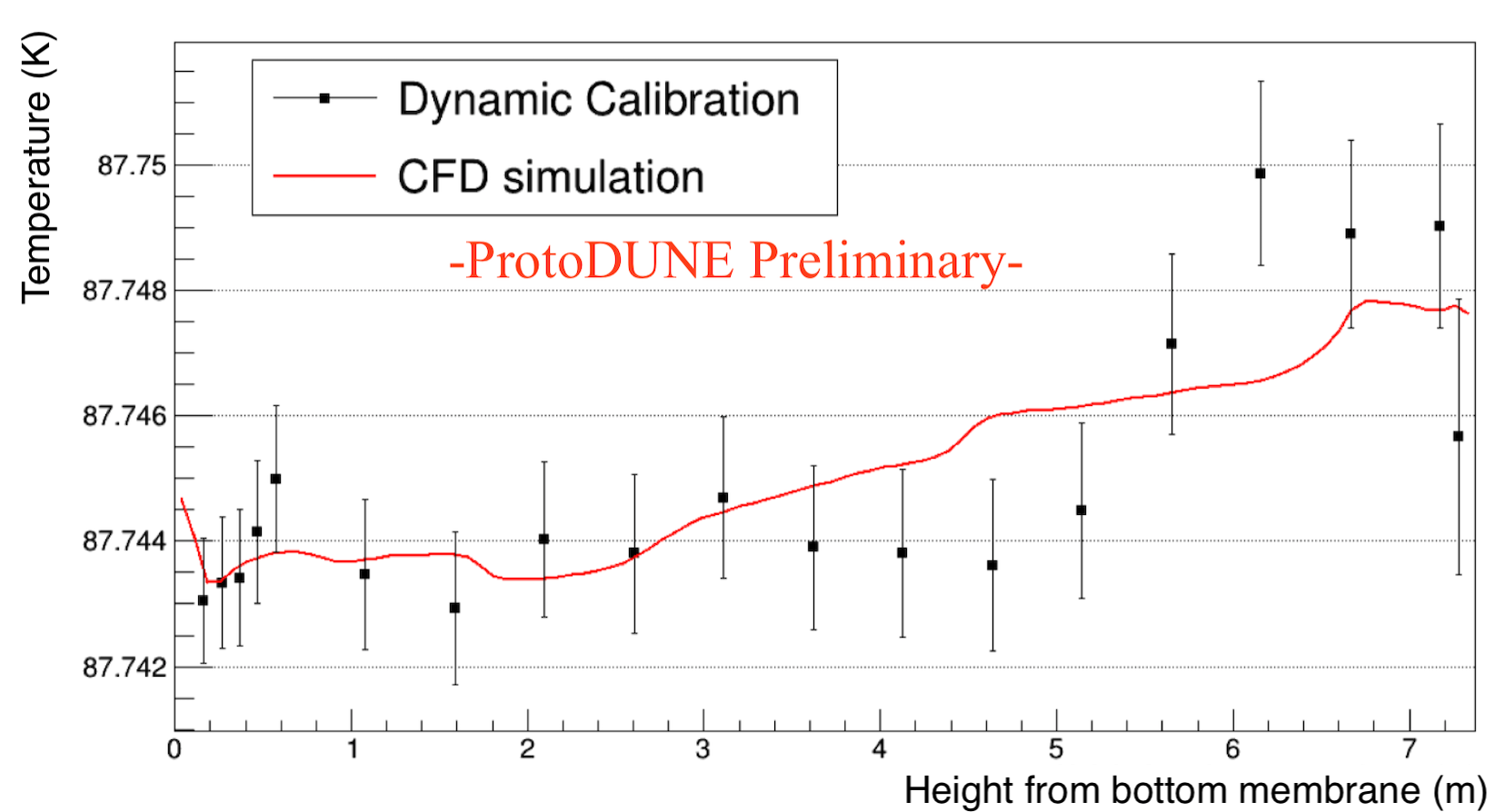}
        \caption{The dynamic temperature profiler sensor values compared to the CFD simulation prediction, as a function of height from the detector bottom.}
        \label{fig:plot_3}
    \end{minipage}%
    \hfill
    \begin{minipage}{0.48\textwidth}
        \centering
        \includegraphics[width=0.9\linewidth, height=0.17\textheight]{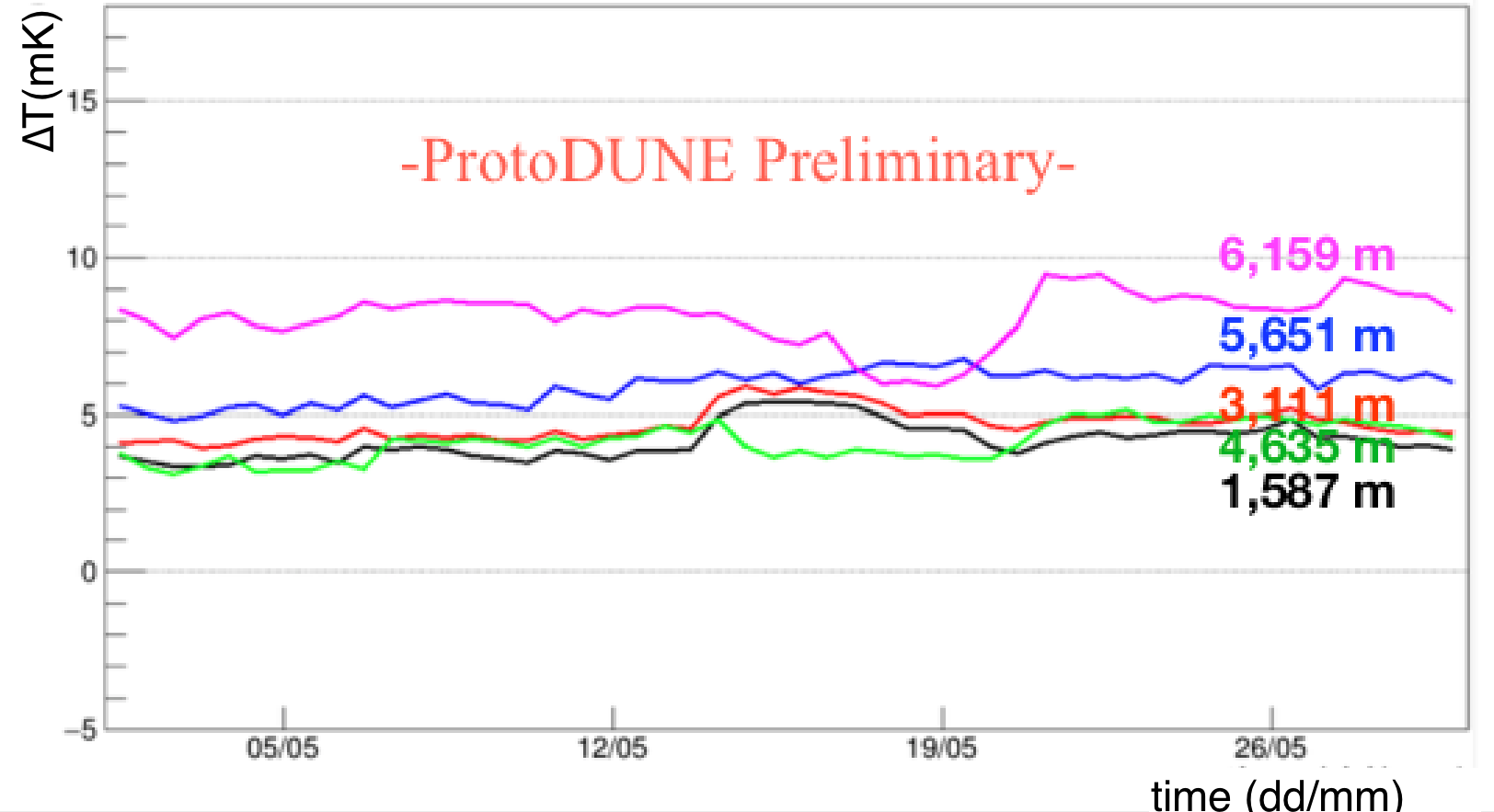}
        \caption{Dynamic temperature profiler sensor values as a function of time. The y-axis values represent the differences in temperature of each sensor with respect to the bottom-most reference sensor.}
        \label{fig:plot4}
    \end{minipage}
\end{figure}

The dynamic temperature profiler was installed during the summer of 2018 and was operational for 2 years. The sensor offsets are consistent over four separate calibration campaigns (Fig.\ref{fig:plot2}). The differences in the offsets are due to local variations during movement and small compared to the overall offset.
Data from the dynamic temperature profile is utilized to cross-calibrate the other (static) temperature sensors in the detector. Finally, all the sensor values are then used to validate Computational Fluid Dynamic (CFD) models \cite{CFD_model} of liquid-argon in the detector. The CFD models will help us to understand argon re-circulation within the detector and its purity (see Fig.\ref{fig:plot_3}).
Stability of the dynamic temperature profiler over two months is shown in Fig. \ref{fig:plot4}. 

\section{Conclusion}
The dynamic temperature profiler was continuously operating in ProtoDUNE since summer of 2018 to summer of 2020. The system was shown to have an outstanding accuracy between 2 and 3 mK and exhibited no fatigue in cryogenic operation.

\bibliographystyle{unsrt}
\bibliography{ms}

\end{document}